\documentstyle[floats,aps,aipbook,psfig]{revtex}

\begin{document}

\title{Low-Energy Spectral Features in GRBs}

\author{Michael S. Briggs}

\address{Department of Physics,
University of Alabama in Huntsville, Huntsville, AL 35899}

\maketitle

\begin{abstract}
I discuss low-energy lines
in gamma-ray bursts.   The process of deconvolving gamma-ray spectral data
and the steps needed to demonstrate the existence of a line
are explained.
Previous observations and the current status of the analysis of the BATSE data
are described.
\end{abstract}

% body of paper here

%%\section*{Introduction}

Spectral lines are highly informative of the conditions of the region in which
they are created.      
Unfortunately, the observational and theoretical understanding of lines
in gamma-ray bursts is confused at best.
Observations made with {\it Ginga} provided strong evidence for
absorption lines, yet BATSE has not  confirmed their existence.
Confirmation of the existence of lines and a better characterization of their
properties might provide the essential clue to the mystery of gamma-ray
bursts.

In this paper 
I will not tell you whether lines exist, instead I will explain what
analysis needs to be done to demonstrate the existence of a line in
a gamma-ray spectrum to assist you in judging the evidence for yourself.
I will briefly review selected observations and the status of
the analysis of BATSE data.
I will concentrate on observational issues
regarding spectral features below 100 keV.
The views expressed are my own.

\section*{Detectors and Gamma-Ray Interactions}

The detection of  gamma-ray lines is much more difficult than
the detection of optical lines, primarily because there is no one-to-one
relationship between the energy of an incident gamma-ray photon and the
energy measured by the detector, which is called the ``energy loss''.
Because of the difference between the energy of
the incident photon and the energy loss,
detected events are referred to as ``counts''.
Secondary problems are the poor signal-to-noise  and signal-to-background
ratios prevalent in gamma-ray astronomy.  
%%These latter problems are
%%reduced in bright gamma-ray bursts because of their great brightness.

\begin{figure}[tb!]
\mbox{
\psfig{figure=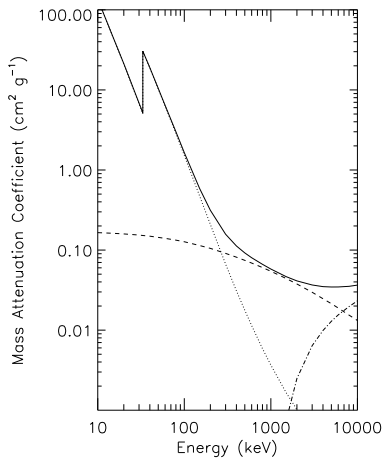,width=50mm,%
bbllx=59bp,bblly=100bp,bburx=552bp,bbury=695bp,clip=}
\hspace{3mm}
\begin{minipage}[b]{63mm}
\caption{
The NaI gamma-ray mass attenuation coefficient $\mu$:
solid line: total; dotted: photoelectric; dashed: Compton; dot-dash:
pair production.    The fractional transmission is $\exp (-{\mu x})$, where
$x$ is the quantity of NaI traversed in g~cm$^{-2}$.
The data are from ref. \protect\cite{Hubbell}. 
\protect\vspace{19mm}  }
\label{fig_cross_sect}
\end{minipage}
}
\end{figure}

The examples in this paper are from data collected with the 
Spectroscopy Detectors (SDs) of BATSE, but the concepts are true for all
gamma-ray detectors---regardless of their energy resolution, the photon
interaction physics is similar.
%%Some detectors, such as Germanium detectors, have superior energy resolution,
%%but the interaction physics is similar.
The SDs are 12.7~cm diameter by 7.6~cm thick crystals of NaI(Tl)
scintillator, each viewed by a photomultiplier tube (PMT) of the same diameter.
When a gamma-ray interacts in the NaI crystal, a fraction of the resulting
ionization energy is converted into scintillation light and measured
by the PMT.
In the energy range of interest, 10 keV to a few MeV, gamma-rays interact
with detector matter primarily by three processes
(Fig.~\ref{fig_cross_sect}):
photoelectric absorption, Compton scattering, and pair production
\cite{Fermi,Hubbell,Knoll}.

\begin{figure}[tb!]
\mbox{
\psfig{figure=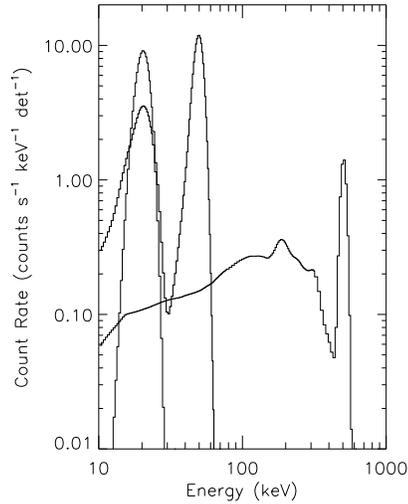,width=55mm,%
bbllx=60bp,bblly=95bp,bburx=550bp,bbury=695bp,clip=}
\hspace{3mm}
\begin{minipage}[b]{58mm}
\caption{
Simulated instrumental energy loss spectra, commonly known as count spectra,
for monoenergetic photons incident on a Spectroscopy Detector,
the detector's surroundings on CGRO and the Earth's atmosphere.
The simulation matches the conditions with which SD 5 observed
GRB~920311 (BATSE trigger 1473), except for clarity the
simulated data extend to energies slightly below where the current
calibration is reliable.
(Data from SD~5 and GRB~920311 are shown in
Fig.~\protect\ref{fig_cont_deconvolve}.)
In each case the input flux is 1~photon
s$^{-1}$~cm$^{-2}$.  
Left curve: photon energy 20~keV; middle, bold curve: 
50~keV; right curve: 500 keV.
\protect\vspace{6mm}  }
\label{fig_delta_inputs}
\end{minipage}
}
\end{figure}

In NaI,  the photoelectric process dominates below $\approx 200$ keV.
In this process 
the gamma-ray is completely absorbed by an atomic electron.  The interaction
is most likely to occur with the inner-most electron  that the gamma-ray
has sufficient energy to ionize.  
Usually, an atomic cascade yields fluorescent X-rays,
which will be photoelectrically
absorbed unless they escape the crystal.
Sometimes an Auger electron is ejected instead.
The left curve of Fig.~\ref{fig_delta_inputs} illustrates a simple case,
the detected energy losses expected from 20 keV photons.   
The fluorescent X-rays have a higher cross-section because of their lower
energy (Fig.~\ref{fig_cross_sect}), so that
they are unlikely to
escape and thus the incident
energy will be totally absorbed, leading to a
simple observed spectrum.
Broadening of the spectral resolution occurs, primarily due to Poisson
fluctuations in the number of photoelectrons produced in the PMT \cite{Knoll}.

A 50 keV photon is also most likely to interact via the photoelectric
process, but the resulting count spectrum is more complicated
(Fig.~\ref{fig_delta_inputs}, bold curve).
The inner-most shell of an iodine atom
accessible to the 20 keV photon of the previous example
is the L-shell.    
The K-shell becomes accessible to photons with energies above 
the K-shell binding energy of 33.17 keV,
resulting in a cross-section increase
(Fig.~\ref{fig_cross_sect}).   A 50 keV photon will typically
ionize a K-shell electron, resulting in a L, M or N to K shell transition 
and an X-ray with an energy
between 28.3 and 33.0 keV \cite{Led}.   Sometimes this photon will
interact by the photoelectric process and the total energy of the initial
photon will be collected, resulting in the full-energy absorption
peak at the right in the bold curve
of Fig.~\ref{fig_delta_inputs}.
However, the fluorescence X-ray is below the K-edge and has a lower
cross-section (Fig.~\ref{fig_cross_sect})
and thus a high probability of escaping
the crystal, resulting in incomplete absorption of the energy of
the incident gamma-ray (the left peak of the
bold curve of Fig.~\ref{fig_delta_inputs}).

Comparing incident photon energies just above and below the 33.17 keV
K-edge of iodine, the higher energy photon has a higher interaction
probability, but more importantly,
a lower probability of complete energy absorption.
For a hard incident spectrum, such as a GRB spectrum, this K-edge 
effect results in a deficit from 33 to $\approx 50$ keV in the count
spectrum, which appears as a 
peak at $\approx 30$ keV
and is predicted by the detector response model
(see Fig.~\ref{fig_cont_deconvolve}).
%% because higher energy photons
%%are less likely to deposit all of their energy and when they fail to do
%%so will produce counts below 33 keV.
This feature has been used as a verification of the calibration of the
SDs \cite{Pac96} and should not be mistaken for an astrophysical line.

At 500 keV, photoelectric absorption occurs only $\approx 20$\% of time,
making a contribution to the full-energy absorption peak
(see Fig.~\ref{fig_delta_inputs}, right curve).
The most likely interaction is Compton scattering,
which transfers only a portion of the incident photon's energy
to an electron.
Complete
energy absorption will occur only if additional interactions occur, either
photoelectric or Compton.
In Compton scattering
the maximum energy transfer to the electron, 331 keV, occurs when the
incident photon is scattered 180$^\circ$.    A range of scattering angles
approaching 180$^\circ$ creates the peak in the curve at 310 keV and the
lack of larger energy transfers in single scattering events causes the
valley above 331 keV (see Fig.~\ref{fig_delta_inputs}).
Correspondingly, the minimum energy of a scattered
photon is 169 keV, so a range of angles approaching 180$^\circ$ for incident
photons that {\it scatter into} the detector
from the spacecraft or the Earth's atmosphere creates the peak at 190 keV.

At yet higher energies  pair production becomes important.

\section*{Spectral Deconvolution and Line Detection}

As described in the previous section, an observed 20 keV count could 
be due to a 20, a 50, or even a 500 keV photon!
%%If one observes 20 keV worth of scintillation light, one does not know
%%whether it came from a 20, a 50, or even a 500 keV photon!
When a single count  is observed, 
it is impossible to
deduce the energy of the incident photon.
If many counts are observed, the incident spectrum can be deduced
within statistical limits in the process known as deconvolution.

We approximate the continuous (as a function of energy) 
detection process with the following discrete 
equation:                          
\begin{equation}
\vec{c} = {\bf D} \vec{p}.
\end{equation}
Here $\vec{c}$ is the vector with the counts versus energy bins.
Normally these data are only available in binned form in order to reduce
telemetry requirements---as long as the energy bin widths are small compared
to the detector's energy resolution essentially no information is lost.
Similarly, $\vec{p}$ is the vector representing the incident photon
spectrum.
%%binned sufficiently finely.
The detector is represented by
${\bf D}$, the detector response matrix, which is obtained via
Monte Carlo simulations of gamma-ray interactions in a computer model
of the detector (e.g., \cite{Pen95}).

The obvious solution is
\begin{equation}
\vec{p} = {\bf D}^{-1} \vec{c}.
\end{equation}
This simple approach
does not work if one wishes to deduce information at a resolution
at or better than the intrinsic resolution of the detector, which is the goal
of analyzing data for the presence of lines.
The problem is that if the energy widths of the rows and columns of ${\bf D}$
are comparable to the detector resolution, then neighboring columns
will be very similar and ${\bf D}$ will be nearly singular.
Inverting ${\bf D}$ will be numerically unstable and the solution $\vec{p}$
will be unreliable, especially in the presence of statistical fluctuations
in the observed count spectrum $\vec{c}_{\rm obs}$.
                                                
\begin{figure}[tb!]
\mbox{             
\psfig{figure=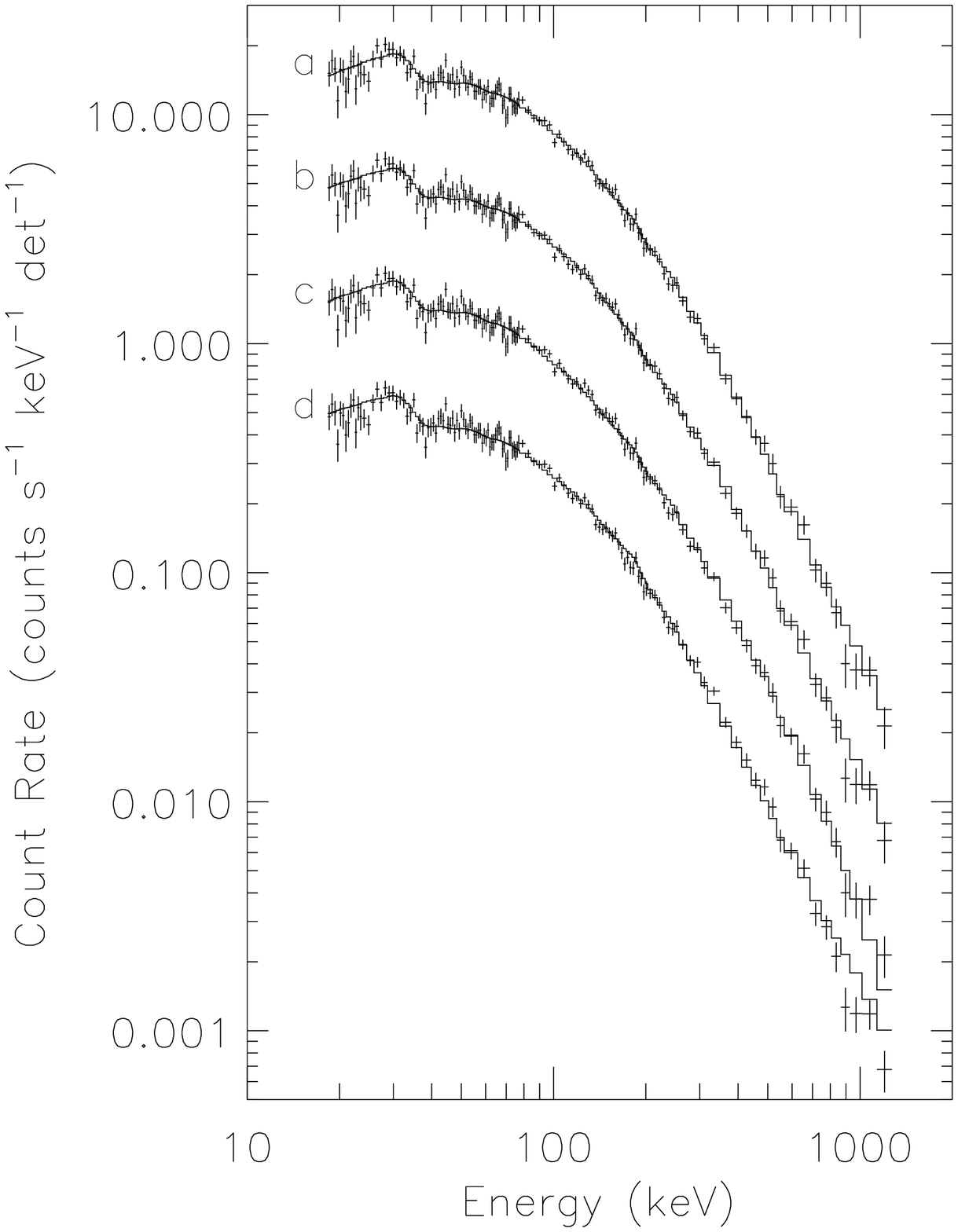,width=58mm,%
bbllx=55bp,bblly=100bp,bburx=525bp,bbury=695bp,clip=}
\psfig{figure=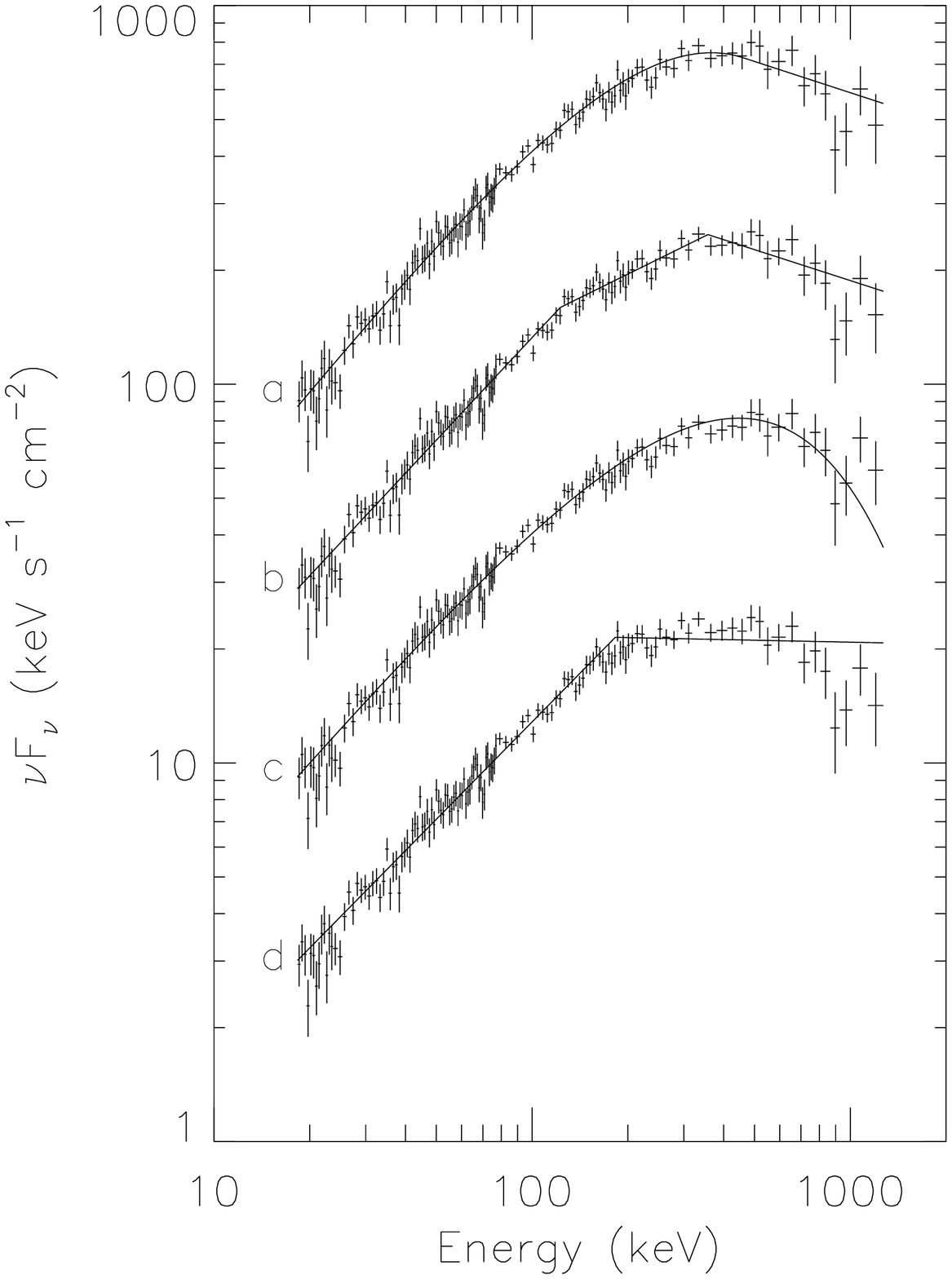,width=58mm,%
bbllx=55bp,bblly=100bp,bburx=525bp,bbury=695bp,clip=}
}
\vspace{0.3mm}
\caption{
To illustrate the photon model dependent nature of the forward-folding
deconvolution method,
identical data
is analyzed using four different photon models. 
Model `a' assumes Band's spectral form \protect\cite{Band93},
obtaining $\chi^2$=223 for 200 degrees-of-freedom (dof).
Model `b' is a power-law with two breaks;  $\chi^2$=227 for 198 dof.
Model `c' is  a power law times an exponential
cutoff; $\chi^2$=244 for 201 dof.  
Model `d' is a power-law with one break; $\chi^2$=267 for 200 dof.
The data are for the interval 1.088 to 26.624~s of GRB~920311 as
observed by SD~5; above 200 keV, the data have been rebinned into wider channels
for display purposes.
Each group after `a' is shifted downwards by $\times \protect\sqrt{10}$.
Left: count data (points) and models (histograms).  
Right: Deconvolved points and models.   
The $\nu F_\nu$ spectrum is $E^2 \times$ the photon flux spectrum.
The photon models
obviously differ---there are also differences in the values of the deconvolved
points,
e.g., the values of right-most point  are $483 \pm 101$, $483 \pm 101$,
$593 \pm 113$ and $449 \pm 98$ keV s$^{-1}$ cm$^{-2}$ 
for models `a', `b', `c' and `d', respectively.
}
\label{fig_cont_deconvolve}
\end{figure}

Instead, the standard approach in astrophysics has become 
that of forward-folding
\cite{Trombka,Fen83,Loredo_Epstein}.
A parameterized spectral model is {\it assumed} and used to
calculate $\vec{p}_{\rm model}$, from which eq. 1 yields a model
count spectrum $\vec{c}_{\rm model}$.  The model count spectrum
$\vec{c}_{\rm model}$ is compared to the observed count spectrum
$\vec{c}_{\rm obs}$ using some statistical measure such as $\chi^2$ or
likelihood and the parameters of the photon model are optimized so as to
minimize the discrepancy between $\vec{c}_{\rm model}$ and $\vec{c}_{\rm obs}$
according to the chosen statistical measure.

It is very important to realize that a solution obtained by forward-folding
is not unique but is rather photon model dependent
\cite{Fen83,Loredo_Epstein}.
Even if a model fit results in a good $\chi^2$ value, another, possibly
unknown, model might result in an equal or better $\chi^2$ value.
This is shown in Fig.~\ref{fig_cont_deconvolve},
where solutions `a' and `b', based upon different photon models,
have essentially identical $\chi^2$ values.
What a forward-folding solution provides is the parameter values of
the {\it assumed} photon model \cite{Loredo_Epstein}.

\begin{figure}[tb!]
\mbox{
\psfig{figure=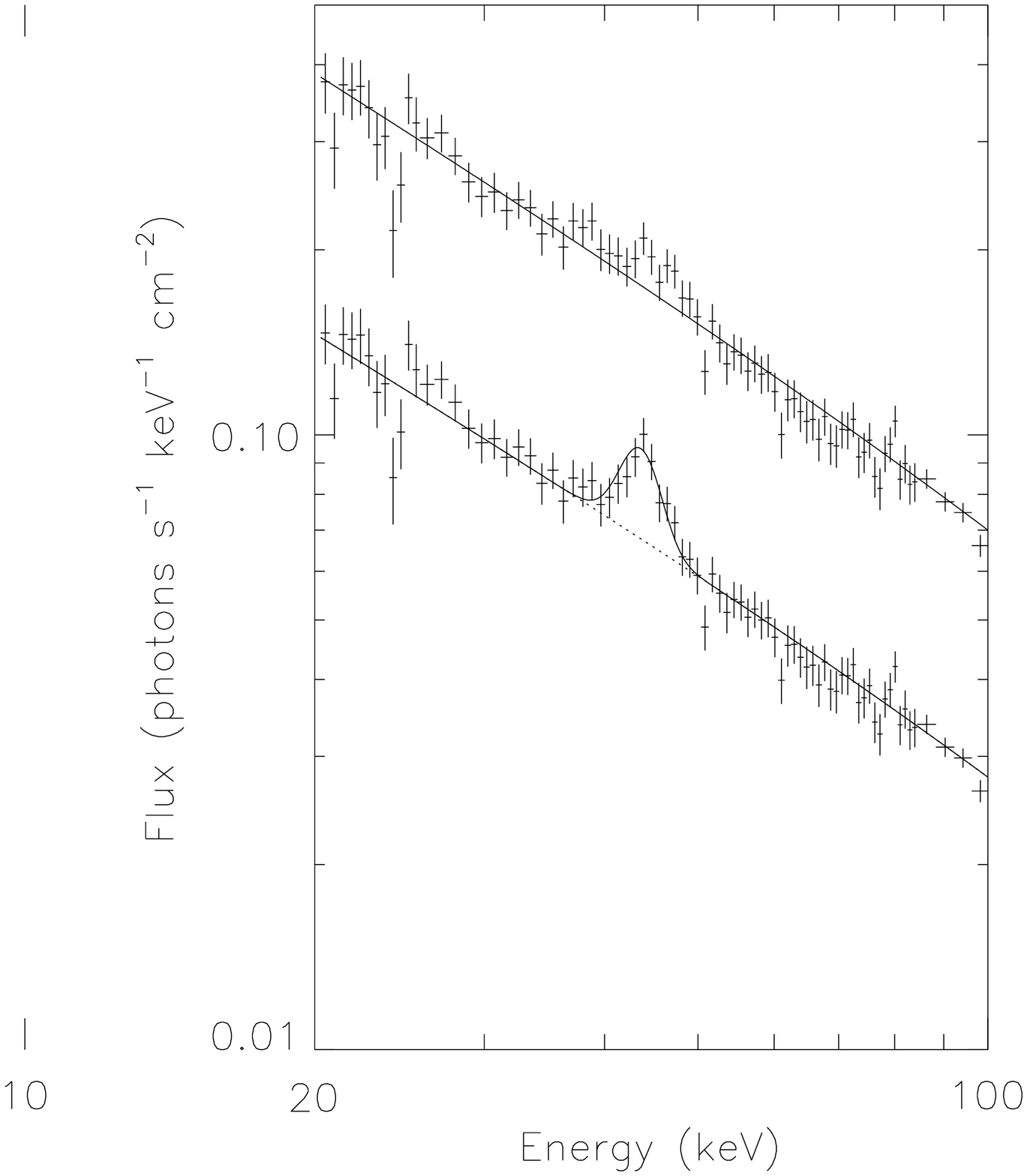,width=52mm,%
bbllx=70bp,bblly=100bp,bburx=540bp,bbury=695bp,clip=}
\hspace{3mm}
\begin{minipage}[b]{56mm}
\caption{
To illustrate the dependence of the deconvolved photon  points on the
model for a model containing a line,
two deconvolutions of the same data are shown.   For the top curve,
a continuum model is assumed, while for the bottom curve (shifted downward
by $\times 2.5$), an additive Gaussian line
is also included in the model.   The data are that of the line candidate in
GRB~940703---for count data and models see ref. \protect\cite{Bri96}.
\protect\vspace{15mm}  }
\label{fig_line_deconvolve}
\end{minipage}
}
\end{figure}

In gamma-ray astrophysics, deconvolved photon points are frequently
obtained by scaling the model photon spectrum by the ratio of the
observed over modeled counts
\cite{Loredo_Epstein}---Fig.~\ref{fig_cont_deconvolve} (right) is an example.
This practice is potentially misleading
because the deconvolved points are model dependent
yet there is an almost irresistible temptation to regard them as 
incident spectrum ``data'' points.
In the examples in Fig.~\ref{fig_cont_deconvolve}, based upon alternative
continuum models, the differences in the deconvolved points are subtle
and the practice not too pernicious.
When a line with width comparable to or smaller than the
detector's intrinsic resolution is
considered,
the deconvolved spectra can exaggerate
the significance of a line---see Fig.~\ref{fig_line_deconvolve}.
Papers analyzing spectra in the X-ray band generally
show a graph of the count data and
model and a graph of the residuals (e.g., \cite{Koyama}).
Only rarely is a graph of a deconvolved spectrum presented.
The gamma-ray community is advised to emulate this practice.

Because of the limitations of gamma-ray spectral data,
the following steps are
necessary to demonstrate that a line feature exists:
\begin{list}{$\bullet$}{\setlength{\itemsep}{0pt} \setlength{\parsep}{0pt}}
\item {\it Deconvolve spectra using the forward-folding technique.}
   This is the standard approach in X-ray and gamma-ray spectroscopy.
\item {\it Show the data and the model in counts rather than deconvolved
        photons.}
       The fit is performed by comparing observed and model counts and
   the data and model should be displayed in this space.
\item {\it Show that the line is 
   significant versus all reasonable continuum models.}
  Suppose that $\Delta \chi^2$ indicates that a line is a statistically
  significant improvement to continuum model A---one might be tempted
   to say that the line
  is ``proven''.  However, if
  the line insignificantly improves continuum model B {\it and} model B 
  has comparable $\chi^2$ to that of A+line, then we cannot say the line
  must exist because continuum model B may be the correct explanation.
   Since we do not know the correct continuum model of GRBs, we must try
  all reasonable ones \cite{Fen83,Fen88}; since all models cannot be tried,
  judgement is needed.
%%
%%  If a line is statistically significant when added to continuum model A, but
%%  insignificant when added to continuum model B which acceptably fits the data,
%%  then we cannot say the line is proven to exist
%%  because continuum model B may be the correct explanation.
%%
  In practice, this means that a sufficiently flexible continuum model should
  be used \cite{Fen88}.   
  With current data, an appropriate choice for all bursts
  is Band's function \cite{Band93,Band_rev},
   which is usually the best-fitting
  model for bright bursts (e.g., Fig.~\ref{fig_cont_deconvolve}).
  A power law with multiple breaks is also sufficiently flexible,
 but has the disadvantage of not being smooth.
   An evolving or two-component continuum spectrum
   might mimic a line when
   a single continuum model is used for the analysis \cite{Hard86,Pen_2cont}.
\item {\it Provide a quantitative evaluation of the statistical significance
   of the feature.}
   Without such an evaluation one does not know the strength of the evidence
   that a line is real rather than a fluctuation.
   It has been traditional to use the F-test (e.g., \cite{Fen88,Barat93,Pal94}),
   but the BATSE team has recently realized that $\Delta \chi^2$ is more
   appropriate for this case, where one knows the uncertainties from Poisson
   statistics \cite{Lamb95,Eadie,Band96}.
   In evaluating the significance of a feature, the number of trials should
   be considered.
\item {\it Consider possible systematic errors.}
    Does the line centroid match that of a background line?  If so, is the
    background subtraction adequate?  Is the detector model and calibration
    adequate?
\item {\it If possible, evaluate the feature for confirmation/consistency.}
    If more than one instrument, or more than one detector of an instrument,
    observed the burst with appropriate energy coverage and resolution, then
    the two observations should be compared.    Ideally, both detectors
    would show the feature to be statistically significant, thereby
    confirming the detection.    Depending on the sensitivity of the detectors
    and the strength of the burst, this may not be possible.   In any case,
    consistency between the data of the two detectors is required.
\end{list}

\section*{Line Observations}

Table 1 summarizes the analyses methods used in previous observations of
gamma-ray burst spectral features below 100 keV
and Table~2 presents the results of those analyses.
Many of the previous analyses were done before the gamma-ray community
 realized the importance of all of 
the steps listed in the previous section and before computer improvements
made the forward-folding approach practical.
Probably due to improvements in the sensitivity of detectors, 
there has been historical progression towards
continuum models with more parameters \cite{Band_rev}.  
Many of the earlier
line analyses were done using continuum models that are now known to be
too simple, raising the possibility that the line detections are an
artifact of the chosen continuum model. 
Unfortunately, these problems mean
that some of the previous analyses
must now be judged as inconclusive with regard to the existence of gamma-ray
burst lines.    Considering the importance of this question, reanalyses
are merited.

\begin{table}[b]
\caption{Low-Energy Spectral Features: Analysis Methods}
\begin{tabular}{cccccc}
Instrument & References     & Forward- & Count Data & Flexible &  Statistical \\
           &                & Folding  & \& Model   & Continuum&  Significance\\
           &                & used?    & shown?     & Model used? & evaluated?\\
\tableline
Konus  & \cite{Maz81,Maz82,Maz83}              &
                               No      &   No       &   No     &    No  \\
HEAO A-4 & \cite{Hue84,Hue87}                  &
                               No      &   No       &   No     &    Yes \\
{\it Ginga} & \cite{Fen88,Gra92,Mur88,Wang,Yoshida}  &
                               Yes     &   Yes      &   Yes    &    Yes \\
Lilas & \cite{Barat93}                         &
                               Yes     &   No       &   Yes    &    Yes \\
\end{tabular}
\end{table}

\begin{table}[t]
\caption{Low-Energy Spectral Features: Results}
\begin{tabular}{cccccc}
Instrument& GRB     & References & $\Delta \chi^2$ &  Centroids & equiv. width\\
          &         &            &                 &   (keV)    &  (keV)    \\
\tableline
Konus  & many\tablenote{Unique amongst the results listed in this table,
 one GRB had an emission feature.}
  & \cite{Maz81,Maz82,Maz83}   &    &  30--70  &  5--15  \\
\tableline
HEAO A-4  & 780325\tablenote{First peak of the burst.} &
\cite{Hue87} & 14.0  &  76$\pm$5  &  24 \\
          & 780325\tablenote{Second peak of the burst.} &
 \cite{Hue87}    & 16.6  &  49$\pm$3  &  13 \\
          & 780608           &  \cite{Hue87}  &       &  66$\pm$7  &  14 \\
\tableline
{\it Ginga}     & 880205  & \cite{Mur88,Fen88}  &  48.5   & 19.7$\pm$0.7,
             38.0$\pm$1.6  &  3.7, 9.1  \\
          & 870303\tablenote{Interval `S1'.} & 
\cite{Gra92}  &  30.6  &  21.1$\pm$1.1  &   10.5  \\
          & 870303\tablenote{Interval `S2'. \protect\vspace{-2mm}} & 
\cite{Mur88,Gra92}  &  26.2  &  21.4$\pm$0.7,
              2$\times$21.4  &  4.8, 8.4  \\
       & 890929 &\cite{Yoshida} & 25.6  & 26.3$\pm$1.5, 46.6$\pm$1.7 & 4, 8 \\
\tableline
Lilas & 890306 & \cite{Barat93} & 89.0 &  11.2$\pm$0.5, 34.6$\pm$1.4 &   \\
\end{tabular}
\end{table}

%%All of the features described in Table~2 are absorption features expect
%%for one 45 keV emission feature reported by Konus \cite{Maz81}.
%%Lines were seen by HEAO in two different peaks of GRB~780325 \cite{Hue87} and by
%%{\it Ginga} in two different intervals (S1 and S2) of GRB~870303 \cite{Gra92}.

The Konus observations pioneered GRB line studies.    The analysis technique was
cleverly designed to minimize the computations required: the count spectra
were deconvolved using a standard model and  the resulting photon spectra
were iteratively improved \cite{Maz83}.    The continuum model used
was optically thin thermal bremsstrahlung, $\propto E^{-1}\exp{-E/E_0}$.
The explanation of the deconvolution procedure shows one count spectrum
as an example \cite{Maz83}.
It has been frequently suggested that the Konus lines
might be harmonics of fundamentals that are below the typical
detector threshold of 30 keV.

The most significant line in the HEAO A-4 data is in the 2{\it nd} peak
of GRB~780325 \cite{Hue87}.
The line significance was evaluated using a simple continuum model, an
exponential, fit to only the data below $200$ keV \cite{Hue87}.
Using a better model and all of the data might raise or lower
the significance.
The deconvolution was done using photons-to-counts efficiencies
derived for an $E^{-2}$ spectrum, a procedure equivalent to 
approximating ${\bf D}$ with a diagonal matrix.
This procedure was justified by the suppression of the Compton scattering
response by the active shielding of the instrument and has the advantage
of conservatively assuming that lines do not exist.
The feature is stated to be visible in two detectors, but only summed
data are shown \cite{Hue87}.

The analysis of the {\it Ginga} data is excellent.
(However, some papers present only deconvolved spectra.)
The chance probability of obtaining a reduction in $\chi^2$ of 48.5 by
adding two lines, assuming that the lines do not exist, is $10^{-8}$!
The harmonic relation between the line centroids is powerful evidence
for the cyclotron resonant scattering interpretation.
A possible concern is that the lines were in the same channels
in all 3 bursts \cite{Fen_pri}---while the centroid energies are
different in GRB~890929, for this event
the gains of the detectors were lower than normal
\cite{Yoshida}.

The recent results of Lilas on GRB~890306 are even more statistically
significant---lines
at 11 and 35 keV are reported in a 68~s interval with an F-test probability
of $2 \times 10^{-13}$ 
that this is a chance result \cite{Barat93}.
The $\chi^2$ value for the model containing the lines is somewhat
high: 43.2 for 26 degrees-of-freedom.
This $\chi^2$ value, along with the location of one line near the
detector threshold and the other near the iodine K-edge, are areas for concern.
It would be desirable to see the count model for these data and also for
a comparable spectrum not containing lines.
The spectral evolution during the 68~s interval, which is 
essentially the entire burst, should also be investigated.

\section*{\protect\vspace{-1mm} BATSE and Lines \protect\vspace{-0.5mm}}

The Spectroscopy Detectors were added to BATSE because of the
Konus line observations.
The pre-launch expectations of the BATSE team were that lines would be
readily found. 
Simulations \cite{Band95} and performance tests \cite{Pac96}
verify the ability of BATSE to detect {\it Ginga}-like lines, with generally
poorer sensitivity than {\it Ginga} for 20 keV lines and better for 40 keV 
lines.
However, no lines have been detected with BATSE \cite{Pal94,Band_VS}.
While disappointing, this lack of line detections by BATSE
is not yet in strong contradiction
with the observations of {\it Ginga} \cite{Pal94,Band_VS,Band_stats}.

BATSE has a number of advantages for line studies:
\begin{list}{$\bullet$}{\setlength{\itemsep}{0pt} \setlength{\parsep}{0pt}}
\item State of the art scintillation detectors: excellent energy resolution
and advanced electronics incorporating active PMT bleeder strings and
baseline restoration to handle large pulses and high count rates,
\item Excellent temporal resolution so that lines can be found on whatever
 time scale they may exist,
\item GRB locations from the LADs to aid analysis of the SD data,
\item Good sensitivity,
\item Extensive performance verification, on the ground and in-orbit
      \cite{Pac96},
\item Multiple detectors and detector types, which enable
   consistency/confirmation studies,
%%\item Most comprehensive line search ever performed \cite{Bri96}.
\end{list}

The BATSE team is therefore confident of the instrument's 
ability to detect lines.
Because of the lack of detections by the initial approach, visually examining
burst spectra for lines \cite{Pal94},
the team has implemented the most thorough line search ever conducted.
This comprehensive computer search examines
essentially all time scales and energy centroids below 100 keV.
The new approach has already identified 8 candidate features with
$\Delta \chi^2 > 20$ \cite{Bri96}.
Currently analysis is in progress to determine the consistency of the
multi-detector data and hopefully to confirm some of the features
with multiple detectors.
Until that work is completed the BATSE team considers these features
to be candidates rather than detections.

\section*{The Future}

There is much more work to be done.
On the interpretational side,
bursts from galactic halo or cosmological distances have intrinsic luminosities
10$^4$ or  more greater than deduced in the days of the galactic disk paradigm.
Consequently, the physics of line creation will be different than
previously envisioned.    This has been examined in the context of
sources in the galactic halo (e.g., \cite{Isenberg}), 
but little work has been
done in the context of cosmological models.
Several of the past line observations merit reanalysis using techniques
developed since they were published.
The analysis of the BATSE data continues in order to determine the
reality of the candidates identified by the comprehensive search.
With eight candidates already identified, and typically several high-gain
SDs viewing a burst, the BATSE team
will be able to determine the consistency of the
data and confirm many of the candidate features
or to demonstrate that there is some
problem.
Data are also currently being collected and analyzed with the scintillation
detectors of the Konus-W instrument \cite{Apt96,Maz96}
and the high energy-resolution
detectors of TGRS \cite{Pal96}.
The near future will bring the launch of HETE \cite{Ricker} and
PGS on Mars-96 \cite{PGS}.

\end{document}